\documentclass{iau}
\usepackage{graphicx}
\usepackage{fancyhdr}
\usepackage{textcomp}
\usepackage[font={scriptsize}]{caption}
\title
{Multiphase ISM in low luminosity radio galaxies: 
       A case study of NGC 708}

\author[Multiphase ISM in early type galaxies: 
       A case study of NGC 708]   
{Sahu, Sheetal K$^1$; Pandey, S K$^1$; Chaware, L$^1$; 
Pandge, M B$^{2,3}$
}
\affiliation{$^1$SOS in Physics and Astrophysics, Pt. Ravishankar Shukla
University, Raipur, India 492010
$^2$Mahatma Basweshswar Mahavidyalaya, Latur, Maharastra, India 413512.\\
$^3$ Dayanand Science College, Latur, Maharastra India.413512.
\\email: {\tt
sheetal.phy14@gmail.com; proskp@gmail.com; chaware.laxmikant@gmail.com}}
\pubyear{2015}
\volume{315}  
\pagerange{XXX}
\jname {From interstellar clouds to star-forming galaxies: \\ universal Processes?}
\editors{P. Jablonkaeds, F. van der Tak, \& P. Andr\'e, eds.}

\begin{document}
\maketitle
%
\begin{abstract} 
We present a multi-wavelength study of a nearby radio loud elliptical galaxy NGC 708, selected from the Bologna B2 sample of radio galaxies. We obtained optical broad band and narrow images from IGO 2m telescope (Pune, India). 
We supplement the multi-wavelength coverage of the observation by using X-ray data from Chandra,  
infrared data from 2MASS, Spitzer and WISE and optical image from DSS and HST. 
In order to investigate properties of interstellar medium, we have generated unsharp-masked, color, residual, quotient, dust extinction, H{$\alpha$} emission maps. From the derived  maps it is evident that cool gas, dust, warm ionized H{$\alpha$} and hot X-ray gas are spatially associated with each other. 
We investigate the inner and outer photometric and kinematic properties of the galaxy using surface brightness profiles.
From X-ray 2d beta model, unsharp masking, surface brightness profiles techniques, it is evident that pair of X-ray cavities are present in this system and which are $\sim$5.6 Kpc away from the central X-ray source.\\
%
\end{abstract} 
  \vspace*{-0.9 cm}
\section{Introduction, Observation and Analysis}
%
%
%
This paper reports multiband study of NGC 708 a radio loud elliptical galaxy selected from B2 ${sample}^{[1][3]}$ for which observational data in Radio, NIR, UV and X-ray band are already available through various data archives.
%
The main goal of this multi wavelength study is to investigate possible inter-correlation among different constituents of ISM and coexistence of multiphase ISM in extra-galactic environment. This work adresses a wide variety of issues related to different physical properties 
of dust in early type galaxies
. In particular, the data is used to 
(1) perform multi-band 
surface photometry and generate composite and multicolor profiles for the galaxies.
(2) use color profiles to study morphology of molecular gas, dust and other faint features. 
(3) obtain dust properties from wavelength-dependent dust  extinction and compare it with the properties of dust in the Milky-way. 
(4) compare the spatial distribution of the gas and dust revealed in diffirent wavelength 
bands. 
(5) study morphology and properties of ionized gas and star forming regions detected in the H{$\alpha$} narrow band images. \\
%
%
%
%
The photometric observations were carried out during June 14-17, 2012 using IGO two ̆meter optical telescope at IUCAA, Pune (India).
The integrated exposure time in B, V, R and H{$\alpha$} bands were 3600s, 2700s, 1800s, 5400s respectively.
Typical values of seeing during our observations were in range of $\sim$1.5. 
 Standard techniques were applied to process the data as described in [4][6]. 
 \begin{table}
 \caption{Basic Information, Dust Extinction and Other Properties.}

 {\scriptsize
\begin{tabular}{|l|l|}\hline 

%
%
 {\bf Object name } &   NGC 708   \\ \hline
%
RA (hh:mm:ss) & $01:52:46.48$  \\ \hline 
DEC (dd:mm:ss)& $+36:09:06.6$   \\ \hline 
%
%
Dust Mass $M_{d(IRAS)}$  & {$5.75$} $\pm$ {$0.03$} ($M_\odot$)   \\ \hline 
Dust Temp. $T_d$  & {$30.39$} $\pm$ {$5.7$} (K) \\ \hline 
$SFR_{FIR}$ & {$1.87$}($M_\odot$/Yr)  \\ \hline
  \end{tabular}
  } 
 %
 %
%
{\scriptsize
 \begin{tabular}{|l|lll|}\hline 
%
{\bf Extinct. Coef.} & {\bf $R_B$}$\pm${\bf $\Delta R_B$ } & {\bf $R_V$ }$\pm${\bf $\Delta R_V$ } & {\bf $R_R$ }$\pm${\bf $\Delta R_R$ } \\ \hline
Milky Way & $4.1$ &  $ 3.1 $ &  $ 2.3 $ \\ \\
NGC 708 &  7.0$\pm$ 1.2 &  6.0$\pm$1.2 & 7.3$\pm$ 1.2 \\ \hline 
%
%
Molecular gas  & {$8.6$} $\pm$ {$0.2$} & & \\ 
mass $M_{H2}$ &  ($M_\odot$) & & \\ \hline %
$SFR_{H\alpha}$ &  {$1.75$}($M_\odot$/Yr) & & \\ \hline 
  \end{tabular}
  } 
%
%
%
%
\end{table}
%
%
%
%
%
 The radial profiles of surface brightness ($\mu$)
, position angle (PA), ellipticity (e) and $B_4$ (boxiness/ diskyness) parameters were generated  in images of all the wavebands using the `ellipse' task available in $IRAF^{[4]}$.
 \vspace*{-0.5 cm}

\section{Results and Conclusions}

(1) Radial profiles of surface brightness ($\mu$), position angle (PA), ellipticity (e) and $B_4$, indicates peculiar feature within its central 10 arcsec region with bump in the surface brightness profile near at 2-4 arcsec. 
(2) The optical colors (i.e. B-V, V-R, B-R) in conjunction with the Infrared 
color maps, indicates existence of cool gas surrounding the nuclear dust lane in the central region of the galaxy.
(3) The derived extinction curve for NGC 708 shows 
significant difference at the R band, as  compared to the extinction curve of the Milky-way. 
(4) The estimated value of $R_V$ for NGC 708 is  6.0$\pm$ 1.2,  
which is considerbly higher than the canonical value of $R_V$ (3.1) for Milky-way. 
%
%
%
(5) We identified three optical knots (two are located in the east and one is in the west from the center, extended within central 3 kpc region of the galaxy) in HST images and it looks a jet like feature in IGO V-band image.
(6) 
Distribution of ionized gas  in NGC 708 revealed by H$\alpha$ emission map is interesting, namely, an angular V shaped morphology having spiral arm like feature within its central 3 kpc region.
(7) The multi-band imaging analysis of NGC 708 reveals a qualitative spatial correspondence among the morphologies of the dust, H$\alpha$, IR, and radio emissions as well as diffuse X-ray emission. This in turn implies  co-existence of multiphase ISM and their physical association in the sample ${galaxy}^{[2][5]}$. 
(8) The diffuse X-ray emission and residual maps  confirm the 
detection of X-ray cavity $pairs^{[7]}$ as has been found in previous studies.
\begin{figure}
\includegraphics[width=40mm,height=38mm]{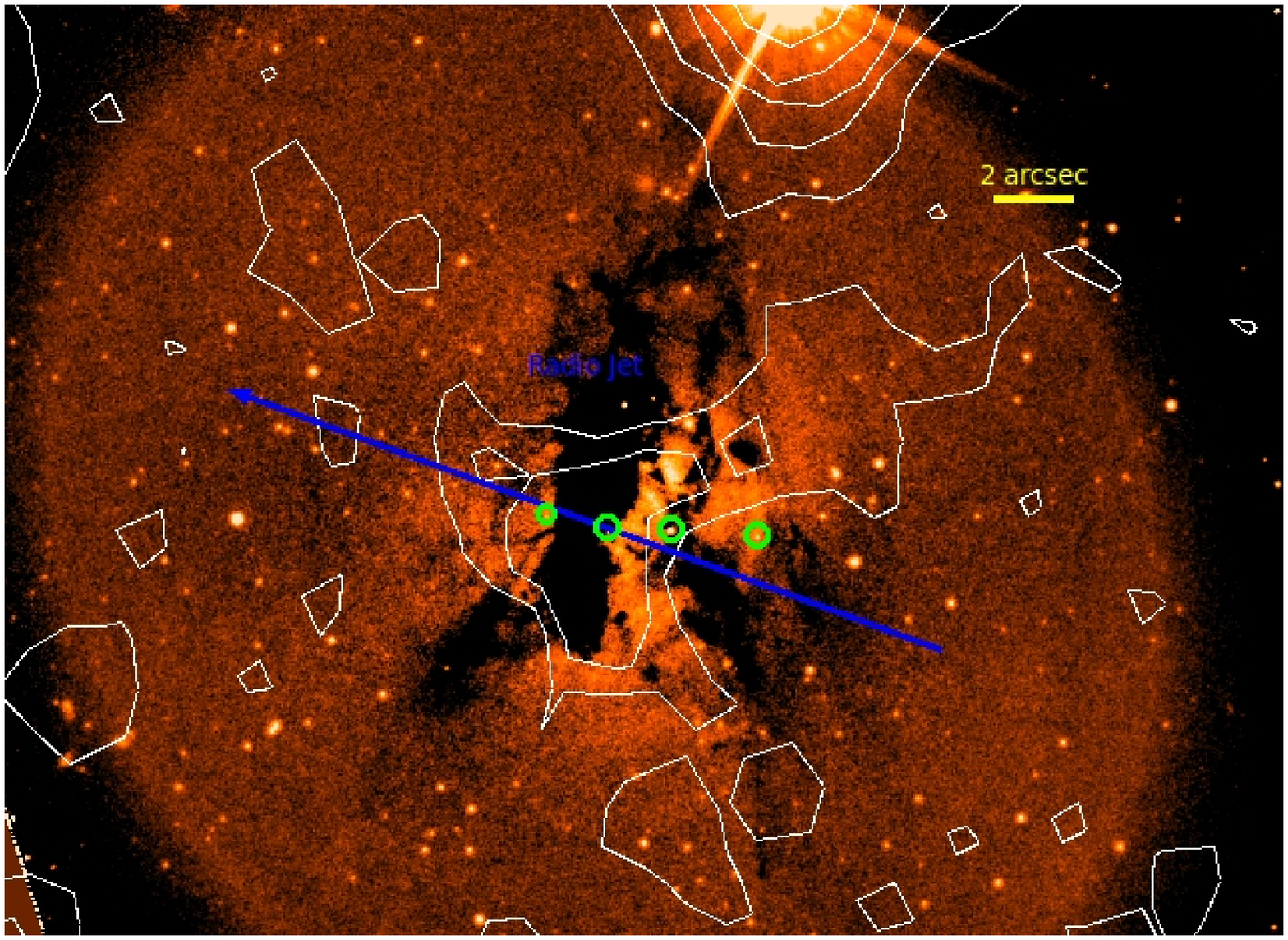}%
\hspace{0.1cm}
\includegraphics[width=40mm,height=38mm]{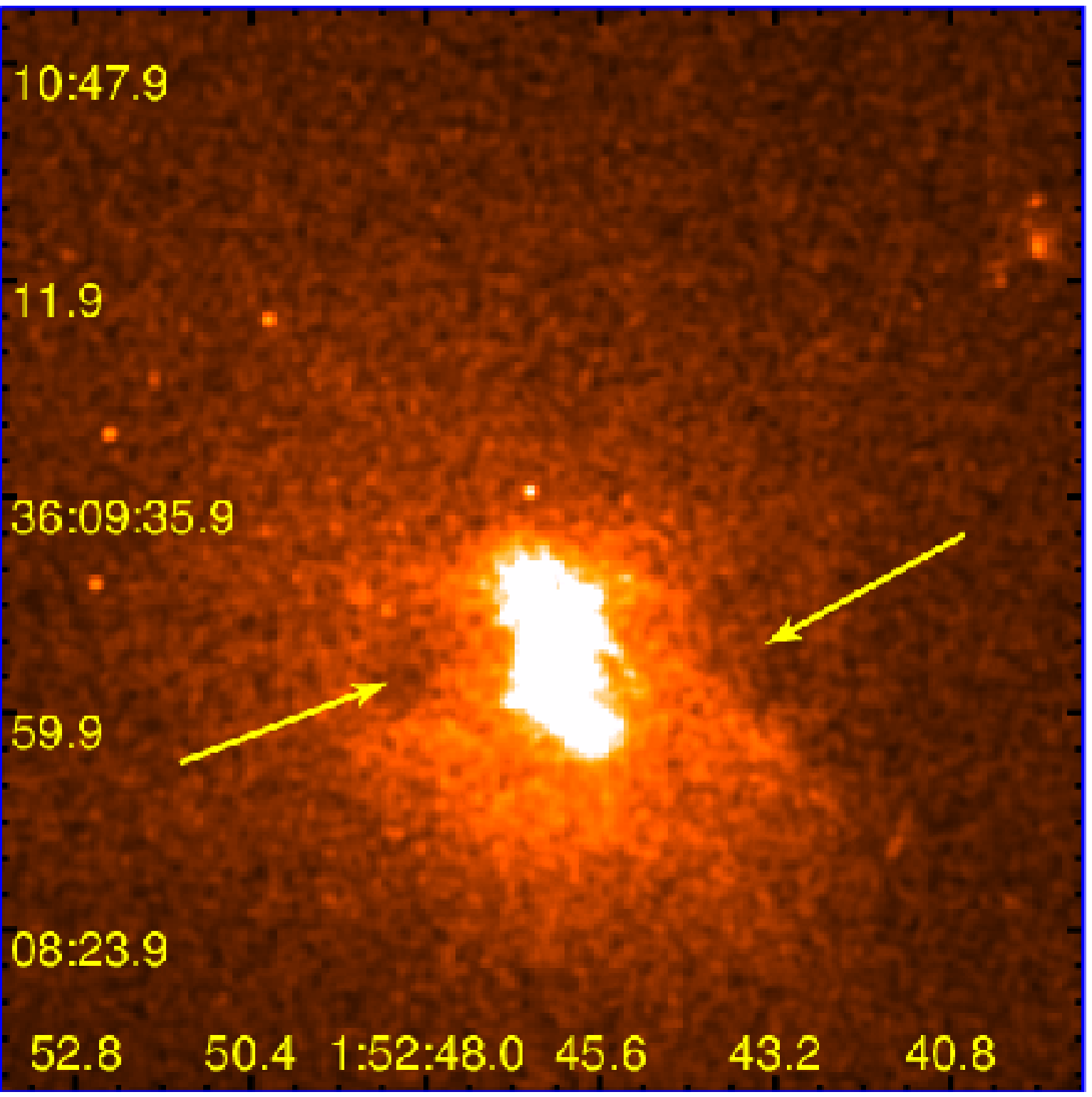}
\includegraphics[width=55mm,height=55mm]{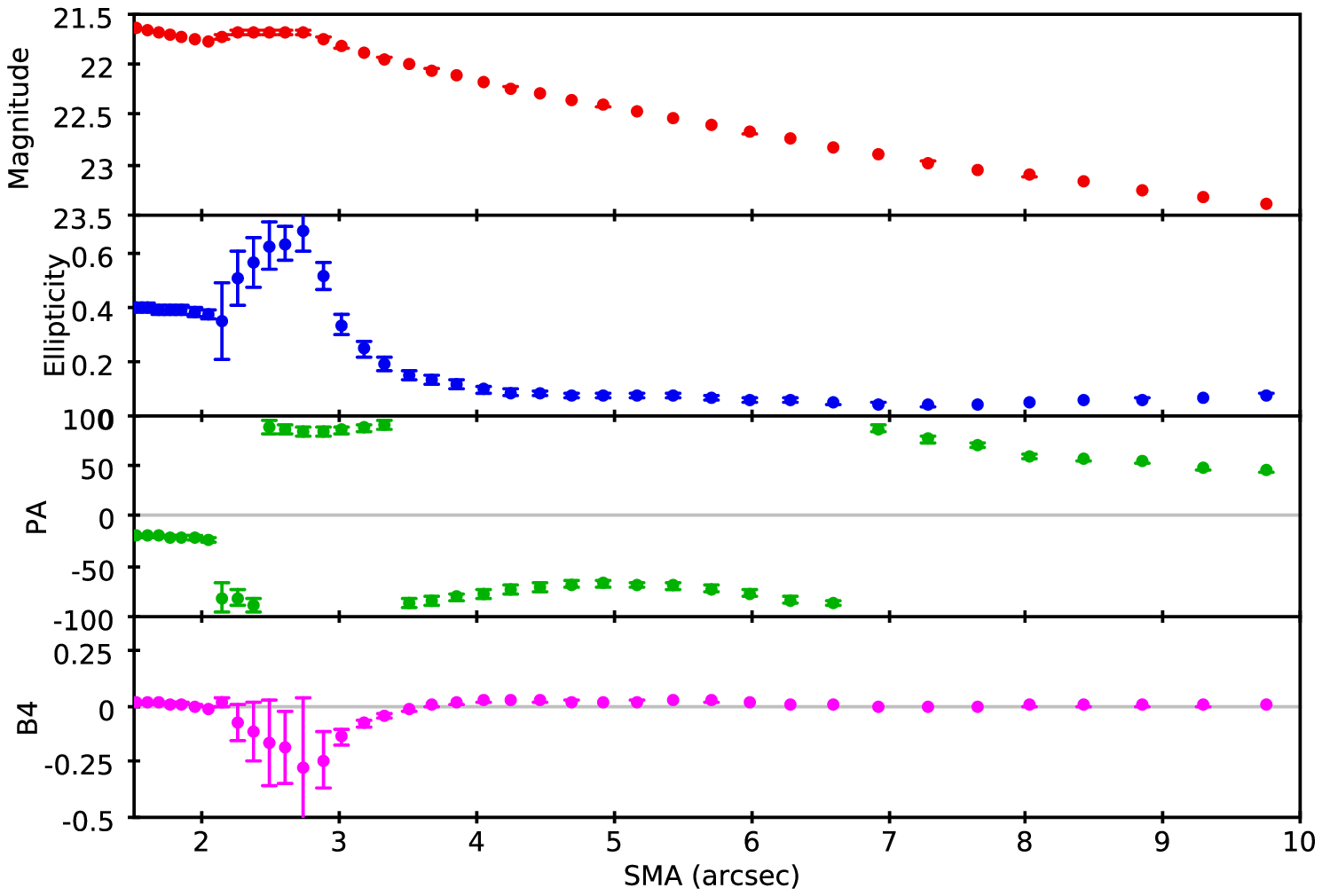}
\caption{{\it Left:}  Residual map {\ttfamily HST[F814W]} showing correlation among dust with radio jet(blue), optical knots (green circles) and IR  ({\ttfamily spitzer[3.6$\mu$]}) emmision (white). {\it Middle:} Residual X-ray (0.3-7.0\,keV 3'$\times$3' \emph{Chandra}) image, reveal substructures in the central region. {\it Right:} Radial profiles of surface brightness ($\mu$), position angle (PA), ellipticity (e) and $B_4$, in V band ({\ttfamily IGO}) within 10 arcsec with bumpy region.
}%
\end{figure}
 \vspace*{-0.7 cm}
\section{References}
%
%
%
[1] Cappetti A, et al, 2000, A$\&$A, 362, 871C.
[2] Finkelman I., et al, 2010, MNRAS, 409, 727F.
[3] Gonzalez-Serrano J.I., et al., 2000, A$\&$AS, 461, 103. 
[4] Jedrzejewski R.I., 1987, MNRAS, 226, 747.
[5] Patil M. K., Pandey S.K., et al, 2007,A$\&$A, 461,103.
[6] Sahu, D. K., Pandey S. K.,et al 1998, A$\&$A, 333, 803.
[7] Pandge M B et al. 2013, A$\&$SS, 345, 183-193.

\end{document}